\documentclass[preprint,prd,noshowpacs,nofootinbib]{revtex4}

\usepackage{doi}
\usepackage{hyperref}
\hypersetup{
  colorlinks=true,        
  linkcolor=blue,         
  citecolor=cyan,         
}

\usepackage{bm}
\usepackage{latexsym}
\usepackage{dcolumn}
\usepackage{amsmath,amsfonts,amssymb}
\usepackage{amsthm}
\usepackage{color}
\usepackage{xcolor}
\usepackage{graphicx,float}
\usepackage{comment}
\usepackage{hyperref}
\usepackage{pdfpages}
\usepackage{epstopdf}
\usepackage[utf8]{inputenc}
\usepackage{subcaption}

\definecolor{green(html/cssgreen)}{rgb}{0.0, 0.5, 0.0}

\def\be {\begin{equation}}
\def\ee {\end{equation}}
\def\bea {\begin{eqnarray}}
\def\eea {\end{eqnarray}}
\def\bc {\begin{center}}
\def\ec {\end{center}}
\def\bfg {\begin{figure}}
\def\efg {\end{figure}}
\def\bi {\begin{itemize}}
\def\ei {\end{itemize}}

\newcommand{\bdm}{\begin{displaymath}}
\newcommand{\edm}{\end{displaymath}}

\begin{document}

\title{ The Mass Gap of the Space-time and its Shape}

\author{Ahmed~Farag~Ali$^{\triangle \nabla}$}
\email{aali29@essex.edu}
\affiliation{\small{$^\triangle$Essex County College, 303 University Ave, Newark, NJ 07102, United States.}}
\affiliation{\small{$^\nabla$Department of Physics, Faculty of Science, Benha University, Benha, 13518, Egypt.}}

\begin{abstract}
\par\noindent
Snyder's quantum space-time which is Lorentz invariant is investigated. It is found that the quanta of space-time have a positive mass that is interpreted as a positive real mass gap of space-time. This mass gap is related to the minimal length of measurement which is provided by Snyder's algebra. Several reasons to consider the space-time quanta as a 24-cell are discussed. Geometric reasons include its self-duality property and its 24 vertices that may represent the standard model of elementary particles. The 24-cell symmetry group is the Weyl/Coxeter group of the $F_4$ group which was found recently to generate the gauge group of the standard model. It is found that 24-cell may provide a geometric interpretation of the mass generation, Avogadro number, color confinement, and the flatness of the observable universe. The phenomenology and consistency with measurements is discussed.

\bigskip
\bigskip
\bigskip
\begin{center}
    ``The knowledge at which geometry aims is knowledge of the eternal''--- Plato.
\end{center}

\end{abstract}


\maketitle

\tableofcontents

\section{Introduction}
\noindent
  In 1947, Snyder established a remarkable step that reconciles the minimal length of measurement with Lorentz symmetry by constructing quantum Lorentzian  space-time \cite{Snyder:1946qz}. The  price was introducing non-commutative geometry and the generalized uncertainty principle (GUP) in Snyder's algebra. For Non-commutative geometry part, it is found to emerge naturally at limits of M/string theory \cite{Connes:1997cr} as higher dimensional corrections of ordinary Yang-Mills theory \cite{Seiberg:1999vs}. Several implications of non-commutative geometry were investigated in quantum field theory and condensed matter systems \cite{Douglas:2001ba,Szabo:2001kg}. For the GUP part, it emerged in several approaches to quantum gravity such as string theory, loop quantum gravity, and quantum geometry \cite{Amati:1988tn,Garay:1994en,Scardigli:1999jh,Konishi:1989wk,Kempf:1994su,Maggiore:1993rv,Capozziello:1999wx}. Phenomenological and experimental implications of the GUP have been investigated in low and high-energy systems  \cite{Das:2008kaa,Pikovski:2011zk,Marin:2013pga,Petruzziello:2020wkd,Kumar:2019bnd,Moradpour:2019wpj,gao2016constraining,Bawaj:2014cda,Girdhar:2020kfl,Ashoorioon:2004vm,Easther:2001fz,dabrowski2019extended,Easther:2001fi}. Useful reviews on quantum space-time and GUP can be found in \cite{Addazi:2021xuf,Hossenfelder:2012jw, Mignemi:2019btr}. Snyder's algebra  is generated by three main generators which are position ${x_\mu}$, momentum ${p_\mu}$ and Lorentz generators ${J_{\mu \nu}=x_\mu p_\nu-x_\nu p_\mu}$. They satisfy the Poincar\'e  commutation relations and suggest new commutation relations that provide a quantum/minimal length as follows:
\be\label{Snyder1}
[x_\mu,x_\nu]=i \hbar ~( \frac{\kappa~ \ell_{Pl}}{\hbar c})^2~ J_{\mu \nu},
\ee
\be\label{Snyder2}
\qquad[x_\mu,p_\nu]=i \hbar~(\eta_{\mu \nu}+(\frac{\kappa~ \ell_{Pl}}{\hbar c})^2~p_\mu p_\nu),\text{with}~~~~~~~~\mu,\nu=0,1,2,3
\ee
where ${\ell_{Pl}}$ is a Planck length, $\kappa $ is a dimensionless parameter that identifies the minimal measurable length, and $\eta_{\mu \nu}=\left(-1,1,1,1\right)$. Eq. (\ref{Snyder1}) introduces the non-commutative geometry and Eq. (\ref{Snyder2}) introduces a GUP.  Both equations are invariant under Lorentz symmetry \cite{Snyder:1946qz}.

\section{space-time quanta and Bekensetin universal bound}
\noindent
In this section, we investigate the physical properties of space-time quanta implied by Snyder's algebra. It is clear that Eq. (\ref{Snyder1}) only vanishes if there is no fundamental minimal/quantum length (i.e $\kappa \ell_{Pl}=0$). This means non-commutative geometry would vanish if there is no minimal/quantum length. On the contrary, we find that the GUP commutation relation in Eq. (\ref{Snyder2})  vanishes. The time-energy commutation relation of Eq. (\ref{Snyder2}) vanishes when:
\begin{eqnarray}
E&=& E_{\kappa}= \pm~ \frac{\hbar~ c}{\kappa~\ell_{\text{Pl}}}, 
\label{real}
\end{eqnarray}
where $E=p_0$ and $E_{\kappa}$ represents the maximum bound on energy. The position-momentum commutation relation Eq. (\ref{Snyder2}) vanishes when:
\begin{eqnarray}
      p&=&p_{\kappa}= \pm~ i~ \sqrt{3}~ \frac{\hbar}{\kappa~ \ell_{\text{Pl}}} \label{complex}
\end{eqnarray}
where $p= \sqrt{p_x^2+p_y^2+p_z^2}$,  and $p_{\kappa}$ represents  momentum bound that are implied by minimal length ($\kappa~\ell_{\text{Pl}}$). This implies ``real'' solutions for energy and ``imaginary'' solution for momentum. To put it another way, time and energy can be known with certainty, implying a complete/perfect description of the space-time quanta in terms of time and energy  that is invariant under Lorentz transformation which is satisfied by Snyder algebra. The quanta of space-time in that sense behave like perfect classical physics at the minimum measurable length scale in terms of energy and time. This may be consistent with the asymptotic freedom property of Yang–Mills theory in which quantum fields behave at a short distance in a very similar way to classical fields \cite{Gross:1973id, Politzer:1973fx}.  Vanishing time-energy uncertainty may shed light on the nature of the wavefunction collapse to form the quanta of space-time. Particle mass has been interpreted as a result of wavefunction collapse  \cite{Ellis:1988uk}. We found recently that vanishing uncertainty could resolve the EPR paradox \cite{Einstein:1935rr} and explain the radii values of Hydrogen atom/nuclei  \cite{Ali:2022jna,farag2022completeness}. \\
\noindent
On another side, Bekenstein found a universal bound \cite{Bekenstein:1980jp, Bekenstein:2004sh, Bekenstein:2000ai} that defines the maximal amount of information that is necessary to \emph{perfectly} and completely describes a physical object up to the quantum level. Bekenstein  universal bound is given by:
\begin{eqnarray}
H \leq \frac{2 \pi~ R~ E}{\hbar~ c} \label{BUB1}
\end{eqnarray}
where $H$ is related to thermodynamic entropy as $S=  k_B H$, where $k_B$ is the Boltzmann constant. $H$ gives the number of bits contained in the quantum states in the sphere with radius $R$ that encloses the physical system, and $E$ is the energy of the physical system. According to Snyder's algebra in Eq. (\ref{Snyder1}) and Eq. (\ref{Snyder2}), the sphere that encloses the space-time quanta is identified by a radius $R=\kappa~\ell_{\text{Pl}}$. Since the space-time quanta implies bound on energy given by Eq. (\ref{real}),  the inequality in Eq. (\ref{BUB1}) turns to be equality for space-time quanta that relates the maximum amount of information $H_\kappa$ to the maximum measurable energy $E_{\kappa}$:
\begin{eqnarray}
E_{\kappa}= \frac{\hbar~c }{\kappa~\ell_{\text{Pl}}} ~\frac{H_{\kappa}}{2 \pi}  \label{BUB2}
\end{eqnarray}
When we compare Eq. (\ref{real}) with Eq. (\ref{BUB2}), we get:
\begin{eqnarray}
H_{\kappa}= 2 ~\pi \label{certinity}
\end{eqnarray}
that completely describes the quanta of space-time. Notice here that $H_{\kappa}$ depends only on $\pi$ and is independent of $\kappa$ and nature constants.

\section{Shape of space-time quanta}
\noindent
A natural question arises, what is the geometric shape of the space-time quanta? To answer this question, we need first to apply the energy-momentum relation that is satisfied by Snyder's algebra \cite{Lu:2011fh}:
\begin{equation}
    E^2-p^2 c^2=m^2c^4 \label{QEPM}
\end{equation}
where $m$ is the mass of the physical object. Substituting Eq. (\ref{real}) and Eq. (\ref{complex}) into Eq. (\ref{QEPM}), we get:
\begin{equation}
    m_\kappa= \frac{2~}{~{\kappa}~ \ell_{\text{Pl}}}  \frac{\hbar}{c}\label{ml}
\end{equation}
This means that the mass of the space-time quanta is completely determined by its unique length and it is strictly a ``positive real value''. It is worth mentioning that generalized forms of Snyder algebra with de Sitter background imply modification of dispersion relation \cite{Banerjee:2011ag} that may introduce corrections to the mass of space-time quanta obtained in Eq.(\ref{ml}). 
The flatness of quantum Lorentzian space-time suggests that the space-time quanta be described by 4-polytope geometry \cite{coxeter1973regular}. We use elementary particle physics as a guide to building the quanta of space-time.  The standard model of particle physics has 25 fundamental particles that include 12 fermions (quarks and leptons), 4 gauge bosons that carry electromagnetic force and weak nuclear force, 8 gluons that carry strong nuclear force, and 1 Higgs scalar field. The quanta of space-time must carry a signature of information from all these fundamental particles that constitute the fundamental structure of nature. The space-time quanta must be ``self-dual'' as well to preserve its uniqueness. Therefore, we look for a highly symmetric 4-dimensional geometric object that is self-dual and identified by only unique length which could represent the fundamental particles of nature on its vertices. This could be a uniform 4-polytope which is a 4-dimensional object with flat sides/faces and is vertex-transitive symmetric which means an isometric map of any vertex onto any other. A regular 4-polytope has the highest degree of symmetry as its faces/cells are regular polytopes and transitive on the symmetries of the polytope. More symmetries are found in regular 4-polytopes that define the convex region as a subset that intersects every line into a single line segment. There are two self-dual convex regular polytopes that are 5-cell, which has 5 vertices, and 24-cell which has 24 vertices. We choose the 24-cell because it has enough vertices to assign with elementary particles. A representation in a plane of a regular 24-cell is given in Fig. (\ref{24-cell-3colors}):
\begin{figure}[H]
    \centering
    \includegraphics[width = 0.5\textwidth]{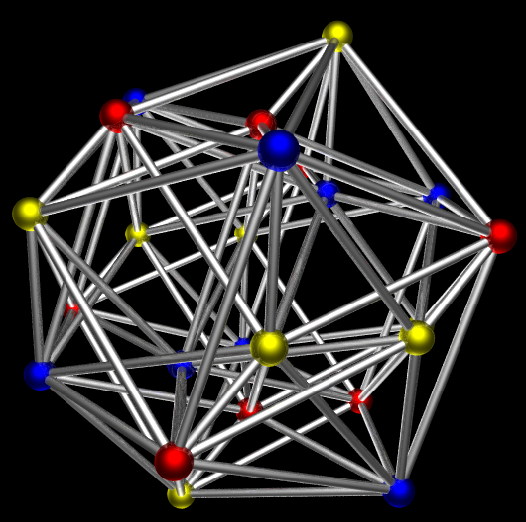}
    \caption\*{24-cell diagram: The vertices of 24-cell are divided into three subsets of three colors, each is a regular 16-cell. Image copied from this \hyperlink{https://commons.wikimedia.org/wiki/File:24-cell_tricolor.png}{ link} with  \hyperlink{https://creativecommons.org/licenses/by-sa/4.0/deed.en}{CC-BY-SA-4.0} license }
    \label{24-cell-3colors}
\end{figure}
\noindent
The 24-cell geometric properties can be summarized as follows:
\begin{itemize}
    \item Its boundary in 3-dimensions forms 24 octahedral cells with six meeting at each vertex, and three at each edge
    \item It has 96 triangular faces, 96 edges, and 24 vertices. The vertex figure is a cube 
    \item self-dual \cite{wenninger2003dual}
    \item identified by one length where edge length equals the distance between the center and vertex (radius)
\end{itemize}
The 24-cell exists in 4- dimensional Euclidean geometry, but Minkowski space-time is 4- dimensional ``pseudo-Euclidean''. This can be simply resolved if time is represented as an imaginary spatial dimension which is well-established in quantum field theory \cite{Wick:1954eu} and in Euclidean quantum gravity \cite{gibbons1993euclidean}. To put it another way, the space-time quanta is represented by a 24-cell with considering time as an imaginary spatial dimension. The covariance principle requires that the space-time quanta should represent the elementary particles of the standard model  \cite{Weinberg:1967tq}. In the next section, we show how to do this representation. \\
\noindent

\section{Symmetry of space-time quanta}
\begin{figure}
     \centering
     \begin{subfigure}[b]{0.3\textwidth}
         \centering
         \includegraphics[width=\textwidth]{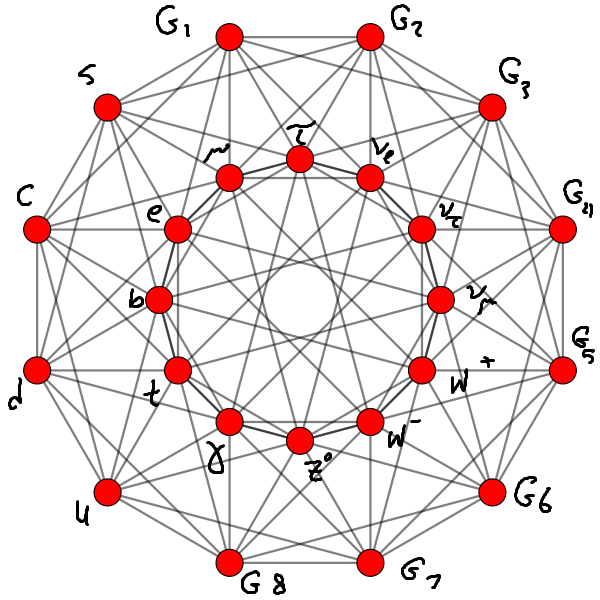}
         \caption{24-cell Coxeter plane that represents elementary particles of the standard model}
         \label{24-cellplane}
     \end{subfigure}
     \hfill
     \begin{subfigure}[b]{0.3\textwidth}
         \centering
         \includegraphics[width=\textwidth]{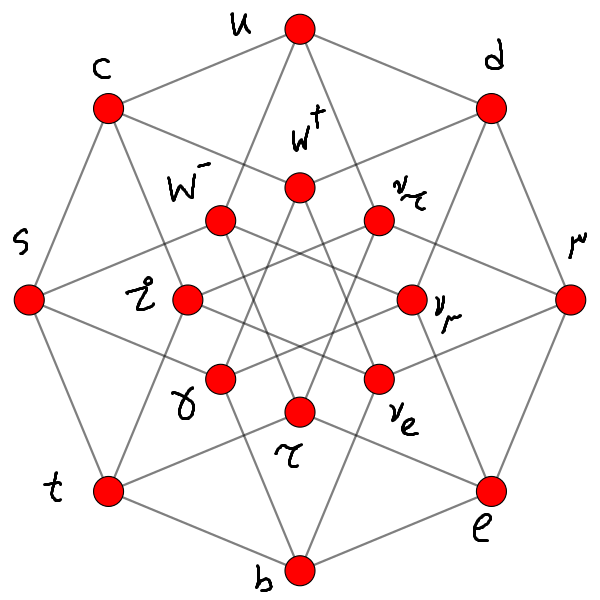}
         \caption{Tesseract Coxeter plane- that act like Higgs particle with 16 vertices assigned to 12 fermions and 4 gauge bosons }
         \label{Tesseract}
     \end{subfigure}
     \hfill
     \begin{subfigure}[b]{0.3\textwidth}
         \centering
         \includegraphics[width=\textwidth]{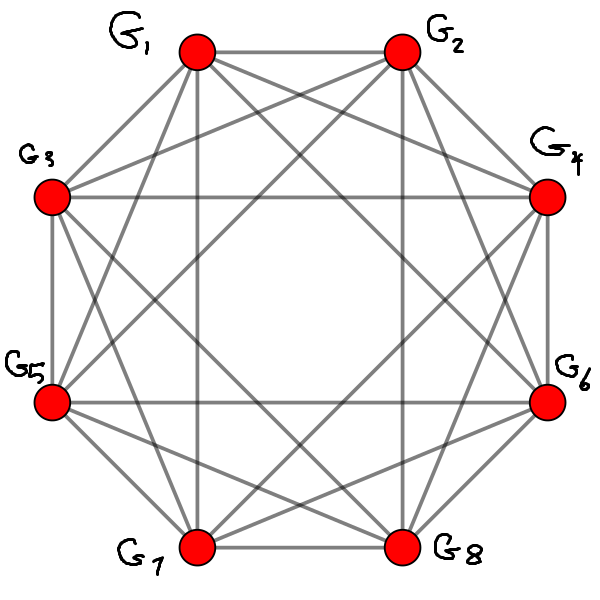}
         \caption{16-cell Coxeter plane that acts like eight gluons that may explain strong force confinement}
         \label{16cell}
     \end{subfigure}
        \caption{representation of fundamental particles on vertices of  24-cell, 16-cell and Tesseract }
        \label{Coxeter plane}
\end{figure}

\noindent
The symmetry group of 24-cell is the  Weyl/Coxeter group of $F_4$ group \cite{coxeter1973regular}. The 48 root vectors of $F_4$ represent the vertices of the 24-cell in two dual configurations. $F_4$ group is one of five exceptional simple Lie groups that are non-abelian and do not have nontrivial connected normal subgroups. The remarkable importance of the $F_4$ group has been realized recently in \cite{Todorov:2018mwd} to explain the gauge symmetry of the standard model of particle physics. It is proved that $F_4$ has two stabilizer groups that are $H_1= (SU(3)\times SU(3))/\mathbb{Z}_3$ and $H_2=Spin(9)$ that their intersection $H_1\cap H_2$ generates the standard model gauge symmetry $(SU(3)×\times SU(2)\times U(1))/\mathbb{Z}_6$. Further details and implications can be found in \cite{ Bernardoni:2007rf,baez2018exceptional,Boyle:2020ctr,Dubois-Violette:2016kzx,Boyle:2019cvm,Todorov:2019hlc,Bhatt:2021cpg,Vaibhav:2021xib,Todorov:2018yvi,Furey:2022yci,Krasnov:2019auj,Dubois-Violette:2018wgs,Furey:2018yyy}. This supports our postulate of identifying the space-time quanta as the 24-cell. In addition, a geometric connection between 24-cell and Calabi-Yau Threefolds with Hodge Numbers (1,1) is realized in \cite{Braun:2011hd} that was useful in determining the mass spectrum of type-IIB flux vacua \cite{Blanco-Pillado:2020wjn}. It is worth mentioning that 24-cell was crystallized to represent the spinfoam topology by trivalent spin network \cite{Aschheim:2012ky}.\\
\noindent
Let us now have a close look at  Fig. (\ref{24-cell-3colors}) which shows a representation in a plane of 24-cell, where the vertices are divided into three subsets of three colors, each being a regular 16-cell. 24-cell vertices can be grouped into three different sets of eight vertices each one defining 16-cells with the rest defining the dual tesseract that has sixteen vertices \cite{coxeter1973regular}.  These three different sets could be used to represent the three colors in quantum chromodynamics. Each set of eight vertices is represented by a 16-cell which is a regular convex 4-polytope that has eight vertices that can represent the eight gluons in the 24-cell. The other sixteen vertices of tesseract could be used to represent the twelve fermions and four gauge bosons. The tesseract is the four-dimensional analog of the cube. In that sense, the 24-cell can be formed by the 16-cell which could be assigned with eight gluons, and the tesseract which could be assigned with twelve fermions and four gauge bosons. Since the Higgs field couples only with the twelve fermions and four gauge bosons,  therefore the tesseract as a whole  could represent the shape of the Higgs particle. In addition, this  property of 24-cell  may explain the geometric meaning of preserving the gluons symmetry which may explain the color confinement. Besides, it explains why gluons could have three colors by grouping 24-cell vertices into three different sets of 16-cell as shown in Fig. (\ref{24-cell-3colors}). It also explains the flatness of the observable universe which is composed of 16 observable particles by the tesseract that has sixteen vertices and which is flat. We can represent the particles by orthogonal projections of 24-cell, 16-cell, and tesseract in the Coxeter plane in two dimensions as in Fig. (\ref{Coxeter plane}).

\noindent

\section{space-time quanta and Spectral mass gap}
\noindent
The solution of the mass gap problem as described in \cite{jaffe2006quantum} requires proving that Yang-Mills theory exists and that the mass of all particles predicted by the theory is strictly positive. Both conditions are satisfied by the space-time quanta. According to Eq. (\ref{ml}), we find that the mass of space-time quanta is a real positive value.
The space-time quanta is quatified by the parameter $\kappa$ in Snyder algebra that was found to generate both non-commutative geometry and GUP at the same time.
For Non-commutative geometry part, it is found to emerge naturally at limits of M/string theory \cite{Connes:1997cr} as higher dimensional corrections of ordinary Yang-Mills theory \cite{Seiberg:1999vs}. Since the space-time is locally flat, the space-time quanta must be described by a 4-polytope. We introduce a geometric and symmetric reason to consider the 24-cell as the space-time quanta. The most important reason is that 24-cell is the Weyl/Coxeter group of  $F_4$ group that can generate the standard model gauge symmetry as shown in recent studies \cite{Todorov:2018mwd}. Therefore, Yang-Mill's theory exists as a $F_4$ group and is explained by the space-time quanta from the first geometric principles as 24-cell. The gluon masses should be related to the 16-cell which is related to the 24-cell we explained in previous sections. We conclude that the space-time quanta introduces a geometric origin of the spectral mass gap \cite{jaffe2006quantum}. The spectral mass gap is entirely determined by the length/radius of 24-cell according to Eq. (\ref{ml}). Recently, it was shown that spectral gaps exist in Hamiltonian with quasicrystalline order  \cite{Hege:2022lkc}. Quasicrystal considerations in Holography, the basic structure of nature, and cosmology are discussed in  \cite{Boyle:2018uiv,aschheim2018starobinsky,irwin2020new,fang2018non,amaral2022exploiting,fang2019empires,irwin2016quantum,Klee}. We think the quantum space-time may be a quasicrystal with a fundamental structure of a 24-cell. Experimental observations of quantum time quasicrystal are reported in \cite{Autti:2017jcw}. This quasicrystal order is expected to follow from simulating Snyder's algebra with considering 24-cell as its fundamental structure. This needs further investigation. 

\section{Phenomenological implications}
\noindent
Phenomenological studies expect to prop information about dark matter at  ILC \cite{Fujii:2017vwa}. We intuit that the fundamental structure of dark matter/energy is the space-time quanta. Previous studies suggested that a minimal length of measurements forms a dark matter candidate \cite{Adler:2001vs}. There is an interesting observation about the 24-cell. All permutations of vertices in 24-cell are given by factorial $24!= 6.2044\times 10^{23}$ which is quite close to the Avogadro number ($6.0221\times 10^{23}$) which determines the ``approximate'' number of nucleons per gram of matter based on thermodynamical and statistical computations. The relative error between the two values is around $2.9\%$. This may shed light on the geometric meaning of the mole unit. The number of observable stars in the universe is between $10^{22}$ to $10^{23}$! \cite{rudnick2003rest,ParticleDataGroup:2020ssz} which is again close to the number predicted by 24-cell. According to the interpretation of creating a mass of space-time quanta by vanishing uncertainty, the minimal length is expected to correspond to the electroweak length scale at which masses are created. Therefore, the minimal length is determined from the measured Higgs mass ($\approx 125.35~ GeV/c^2$) \cite{ParticleDataGroup:2020ssz} and substituting it in Eq. (\ref{ml}) that implies a minimal length as follows:
\begin{equation}
    \kappa~\ell_{\text{Pl}}\approx  10^{-18} ~~\text{m} \label{minimal}
\end{equation}
This would give the length scale at which uncertainty vanishes to form a mass of space-time quanta. In that sense, we reinterpret the Higgs mechanism as a state in which the uncertainty vanishes to form the mass of space-time quanta. The electroweak length scale determines the unique length of the 24-cell. The effective minimal length has been interpreted recently as a charge radius of scattering for every physical particle \cite{Ali:2022ckm}. Recent studies show that the minimum measurable charge radius measured so far is around $10^{-18}$ for the neutrinos,  Higgs, W-bosons, and Z-boson, \cite{Hutauruk:2020mhl,lehnert2014mass,ParticleDataGroup:2020ssz} which is consistent with our interpretations. 

\noindent One notice that the permutations of the 16-cell that forms the gluons part of the 24-cell are equal to factorial $8!$ and the permutation of tesseract that represents  Higgs, photon, W, Z,  quarks, and leptons is given by factorial $16!$.  Based on our model, the 16-cell represents the Planck scale and the tesseract represents the electroweak scale.  The multiplications of their permutations ($8! 16!= 8.4360689\times 10^{17}$) may explain the scale ratio between the Planck scale and the electroweak scale. The measured scale ratio between the Planck scale and the electroweak scale is approximately $ 10^{17}$. This may imply a quantum computational explanation of the hierarchy in nature.

\section{conclusion}
\noindent
We found that the GUP implied by Snyder's algebra vanishes at a specific energy scale. We define this energy scale as the scale of space-time quanta at which wavefunction collapses to form a mass. The mass of space-time quanta forms a mass gap of space-time. The covariance principle requires the space-time quanta to be a 4-dimensional object and to represent the elementary particles. Based on the geometric and symmetric analysis, we propose that the space-time quanta be represented by the 24-cell. First, it is highly-symmetric convex regular 4-polytope and self-dual. Second, The symmetry group of 24-cell is the Weyl/Coxeter group of $F_4$ group that generates the gauge group of the standard model by the intersection of its two stabilizer groups. In addition, the 24-cell has a beautiful geometric property in which its vertices can be grouped into 3 different sets of eight vertices, each defining 16-cell with the rest defining the dual tesseract with 16 vertices. Therefore, we represent 8 vertices of 16-cell with the 8 gluons that may give a geometric interpretation of the color confinement. We represent 16 vertices of tesseract with the 12 fermions and 4 gauge bosons which may explain the flatness of the observable universe. The Higgs particle is represented as the tesseract which explains why Higgs only couple with 12 fermions and 4 gauge bosons and does not couple with 8 gluons. Vanishing uncertainty implies the creation of a mass of space-time quanta, so the length of 24-cell is identified by the electroweak length scale $10^{-18}$ m of mass creation and which is consistent with experimental measurements of the smallest measured value of charge radius of scattering for neutrinos, Higgs, Z-boson, and W boson. We think that a 24-cell symmetry group which is a solvable group of order ``1152'' could be useful in quantum computing. We hope to report on this application in the future. 

\section*{\textcolor{green(html/cssgreen)}{Acknowledgments}}
\noindent \textcolor{green(html/cssgreen)}{I thank the Editor and referee for their meticulous review of this manuscript and express gratitude towards Klee Irwin, Raymond Aschheim, Fang Fang, Richard Clawson, and Dugan Hammock for their enlightening discussions on Polytopes and Quasicrystals. Special thanks to Ahmed Almheiri for his engaging lecture on the Path integral for chords. \\\\
This manuscript is a tribute to my father, Farag Mohamed Ali, whose legacy illuminates my journey, and in memory of Mustafa Shalaby, whose teachings have profoundly inspired my exploration of physics' core principles.}

\bibliographystyle{apsrev4-1}
\bibliography{ref.bib}{}
\end{document}